# The upper critical field in superconducting $MgB_2$

K.-H. Müller, G. Fuchs, A. Handstein, K. Nenkov, V.N. Narozhnyi, D. Eckert
*Institut für Festkörper- und Werkstofforschung Dresden, Postfach 270116, D-01171 Dresden, Germany*

**Abstract.**

The upper critical field $H_{c2}(T)$ of sintered pellets of the recently discovered $MgB_2$ superconductor was investigated by transport, *ac* susceptibility and *dc* magnetization measurements in magnetic fields up to 16 T covering a temperature range between $T_c \sim 39$ K and $T = 3$ K $\sim 0.1 T_c$. The $H_{c2}$ data from *ac* susceptibility are consistent with resistance data and represent the upper critical field of the major fraction of the investigated sample which increases up to $H_{c2}(0) = 13$ T at $T = 0$ corresponding to a coherence length of $\xi_o = 5.0$ nm. A small fraction of the sample exhibits higher upper critical fields which were measured both resistively and by *dc* magnetization measurements. The temperature dependence of the upper critical field, $H_{c2}(T)$, shows a positive curvature near $T_c$ and at intermediate temperatures indicating that $MgB_2$ is in the clean limit. The $H_{c2}(T)$ dependence can be described within a broad temperature region $0.3 T_c < T \leq T_c$ by a simple empirical expression $H_{c2}(T) \propto (1-T/T_c)^{1+\alpha}$, where the parameter $\alpha$ specifies the positive curvature of $H_{c2}(T)$. This positive curvature of $H_{c2}(T)$ is similar to that found for the borocarbides $YNi_2B_2C$ and $LuNi_2B_2C$.

The recent discovery of superconductivity in $MgB_2$ [1] at temperatures as high as 40 K has stimulated considerable interest in this system. $MgB_2$, which has a hexagonal $AlB_2$ structure, is a type II-superconductor. A significant boron isotope effect was observed [2] which is an indication for electron-phonon mediated superconductivity in this compound. Magnetic parameters as the Ginsburg-Landau parameter $\kappa = 26$ [3] and the temperature dependence of the upper critical field $H_{c2}(T)$ [3-6] were determined from transport and magnetization measurements [3-7]. So far, a complete $H_{c2}(T)$ curve was reported for a $MgB_2$ wire sample showing a high residual resistivity ratio of about 25 [7].

In the present paper, the temperature dependence of the upper critical field of a sintered $MgB_2$ pellet was studied in magnetic fields up to 16 T in order to analyse the shape of $H_{c2}(T)$ in the whole temperature range for a sample with a moderate residual resistivity ratio. Polycrystalline samples of $MgB_2$ were prepared by a conventional solid state reaction. A stoichiometric mixture of Mg and B was pressed into pellets. These pellets were wrapped in a Ta foil and sealed in a quartz vial. The samples were sintered at 950°C for two hours. Electrical resistance and the superconducting transition of a sample 5 mm in length with a cross-section of about 1 mm$^2$ (cut from the initially prepared pellet) were investigated in magnetic fields up to 16 T using the standard four probe method and current densities between 0.2 and 1 A/cm$^2$. *AC* susceptibility and *dc* magnetization measurements were performed on other pieces from the same pellet in magnetic fields up to 9 T and 5 T, respectively.

In Fig. 1a, the temperature dependence of the electrical resistance of the investigated sample is shown. The resistivity at 40 K and 300 K are about 6.4 μΩcm and 29 μΩcm, respectively, resulting in a residual resistance ratio (RRR) of approximately 4.5. The midpoint value of the normal-state resistivity of the superconducting transition at zero-magnetic field is 38.8 K. A similar $T_c$ value of $T_c$=39,0 K was determined from *ac* susceptibility data using the onset temperature of the superconducting transition (see Fig. 1b). The field dependence of the electrical resistance of the same sample is shown in Fig. 2 for several temperatures between 36 and 2.9 K. A considerable broadening of the transition curves is observed at low temperatures which may be caused by flux-flow effects at high magnetic fields. The transition widths gradually broaden from 0.2 T at 36 K to 4.5 T at 12 K. A corresponding broadening is found also for resistance-vs.-temperature transition curves at high magnetic fields. In order to compare the transition widths of the two sets of transition curves, the field values $H_{10}$, $H_{50}$ and $H_{90}$ defined at 10%, 50% and 90% of the normal-state resistance are plotted in Fig. 3 as function of the temperature. It is clearly seen that the two data sets determined from field- and from temperature-dependent measurements coincide. Additionally, Fig. 3 shows upper critical field data determined from *dc* magnetization and from *ac* susceptibility measurements. The onset of superconductivity was used to define $H_{c2}$ from *ac* susceptibility. An example for the determination of $H_{c2}$ from *dc* magnetization is shown in Fig. 4, where magnetization data are plotted for two temperatures in an expanded view. The large resolution allows to visualize not only $H_{c2}$, but also the irreversibility field $H_{irr}$ and a large region between $H_{c2}$ and $H_{irr}$ in which the change of magnetization is reversible. The comparison of the upper critical fields obtained from *dc* magnetization, *ac* susceptibility and resistance measurements in Fig. 3 shows clearly, that $H_{c2}^{mag}$ ($H_{c2}$ from magnetization) coincides with $H_{90}$, whereas $H_{c2}^{sus}$ ($H_{c2}$ from susceptibility) agrees approximately with $H_{10}$. The difference between $H_{c2}^{mag}$ and $H_{c2}^{sus}$ can be explained by the inhomogeneity of the sample. It seems that the major part of the sample has the reduced upper critical fields measured by *ac* susceptibility, whereas only a relatively small fraction of the sample shows higher $H_{c2}$ values. One has to take into account that already a narrow current path through the sample with improved parameters is sufficient to produce the observed resistive-transition data. The properties of this small fraction can be detected and systematically investigated by sensitive magnetization measurements. The extrapolation of $H_{10}(T)$ to $T = 0$ yields an upper critical field of $H_{c2}(0) \sim 13$ T for the major fraction of the sample, whereas for the small fraction with improved parameters, $H_{c2}(0) \sim 18$ T is estimated by extrapolation of $H_{90}(T)$ to $T = 0$. Using these $H_{c2}(0)$ values, the coherence lengths $\xi_o = [\phi_o /(2\pi H_{c2}[0])]^{0.5}$ are found to be 5.0 nm (major fraction) and 4.2 nm (small fraction).

A peculiarity of the $H_{c2}(T)$ dependence shown in Fig. 3 is its pronounced positive curvature near $T_c$. Such a positive curvature of $H_{c2}(T)$ near $T_c$ is a typical feature observed for the non-magnetic rare-earth nickel borocarbides $R$Ni$_2$B$_2$C ($R$=Y,Lu) and can be explained by taking into account the dispersion of the Fermi velocity using an effective two-band model for superconductors in the clean limit [8]. We conclude that also our MgB$_2$ samples are within the clean limit in spite of the rather moderate RRR value of 4.5.

The $H_{c2}(T)$ curves in Fig 3 can be described, in a wide temperature range $0.3T_c < T < T_c$, by the simple expression

$$H_{c2} = H_{c2}*(1-T/T_c)^{1+\alpha}, \qquad (1)$$

where $H_{c2}*$ and $\alpha$ are fitting parameter. Notice that $H_{c2}*$ differs from the true value of $H_{c2}(0)$ due to the negative curvature of the $H_{c2}$-vs.-$T$ curve observed at low temperatures. The fit curves in Fig. 3 describing the $H_{10}(T)$ and $H_{90}(T)$ data between 12 K and $T_c$ correspond to values of $\alpha = 0.25$ and $\alpha = 0.32$, respectively. Similar values for the parameter $\alpha$ describing the positive curvature of $H_{c2}(T)$ are known from the rare-earth nickel borocarbides YNi$_2$B$_2$C and LuNi$_2$B$_2$C [9].

It is interesting to note that the $H_{c2}(T)$ curve reported for a high quality MgB$_2$ wire with a RRR value of about 25 showed an almost linear temperature dependence in an extended temperature range: In this case, a positive curvature was observed only in a narrow temperature range near $T_c$. It is also remarkably, that the width of the resistive superconducting transition of this wire at low temperatures is similar to that of our sintered sample. In particular, at T =1.5 K onset and completion of superconductivity (corresponding in our notation approximately to $H_{90}$ and $H_{10}$, respectively) were reported at 16.2 T and 13 T, respectively.

In conclusion, the upper critical field of MgB$_2$ was investigated in a wide temperature range between 3 K and $T_c$. The onset of the superconducting transition of *ac* susceptibility measurements was found to agree with $H_{c2}$ data measured resistively at 10% of the normal state resistance and represents the upper critical field of the major fraction of the investigated sample. A small fraction of the sample exhibits higher upper critical fields which were measured both resistively and by sensitive *dc* magnetization measurements. The investigated sample shows a variation of the upper critical field at T = 0 between $H_{c2}(0)$ = 13 T and about 18 T. The considerable broadening of the resistive transitions at high magnetic fields may be caused partly by the sample inhomogeneity, but it also indicates strong flux-flow effects at high fields. A significant positive curvature observed for $H_{c2}(T)$ in a wide temperature region $0.3T_c < T \leq T_c$ suggest that the investigated $MgB_2$ sample is within the clean limit.

**References**

[1] J. Akimitsu, Symposium on Transition Metal Oxides, Sendai, January 10, 2001; J. Nagamatsu, N. Nakagawa, T. Muranaka, Y. Zenitani, and J. Akimitsu, to be published in Nature.
[2] S.L. Bud'ko, G. Lapertot, C. Petrovic, C.E. Cunningham, N. Anderson, and P.C. Canfield, to be published in Phys. Rev. Lett.
[3] D.K. Finnemore, J.E. Ostenson, S.L. Bud'ko, G. Lapertot, P.C. Canfield, cond-mat/0102114.
[4] Y. Takono, H. Takeya, H. Fujii, H. Kumakura, T. Hatano and K. Togano, cond-mat/0102167.
[5] P.C. Canfield, D.K. Finnemore, S.L. Bud'ko, J.E. Ostenson, G. Lapertot, C.E. Cunningham, and C. Petrovic, cond-mat/0102289.
[6] D.C. Larbalestier, M. Rikel, L.D. Cooley, A.A. Polyanskii, J.Y. Jiang, S. Patnaik, X.Y. Cai, D.M. Feldman, A. Gurevich, A.A. Squitier, M.T. Naus, C.B. Eom, E.E. Hellstrom, R.J. Cava, K.A. Regan, N. Rogado, M. A. Hayward, T. He, J.S. Slusky, P. Khalifah, K. Inumaru, and M. Hass, cond-mat/0102216.
[7] S.L. Bud'ko, C. Petrovic, G. Lapertot, C.E. Cunningham, P.C. Canfield, M.-H. Jung and A.H. Lacerda, cond.-mat/0102413.
[8] S.V. Shulga, S.-L. Drechsler, G. Fuchs, K.-H. Müller, K. Winzer, M. Heinecke, K. Krug, K. *Phys. Rev. Lett.* **80**, (1998) 1730.
[9] J. Freudenberger, S.-L. Drechsler, G. Fuchs, A. Kreyssig, K. Nenkov, S.V. Shulga, K.-H. Müller, L. Schultz, Physica C **306**, (1998) 1.

**Figure captions**

**Fig. 1**

Temperature dependence of (a) resistance and (b) *ac* susceptibility of a polycrystalline $MgB_2$ sample in zero applied field.

**Fig. 2**

Field dependence of the resistance of the $MgB_2$ sample of Fig. 1 for several temperatures between 36 and 2.9 K

**Fig. 3**

Temperature dependence of the upper critical field determined from *dc* magnetization (▼), *ac* susceptibility (Δ) and resistivity measurements with data from field dependence (●) and temperature dependence (o) determined at 10% ($H_{10}$), 50% ($H_{50}$) and 90% ($H_{90}$) of the normal state resistance. Dotted lines are calculated using Eqn. (1) by fitting $H_{c2}^*$ and $\alpha$ to the experimental data.

**Fig. 4**

Expanded view of magnetization vs. applied field for two temperatures. Arrows mark the upper critical field $H_{c2}$ and the irreversibility field $H_{irr}$ for the data measured at T = 27.5 K. Arrows are also used to show how the applied field was changed in the different branches of the magnetization loop.

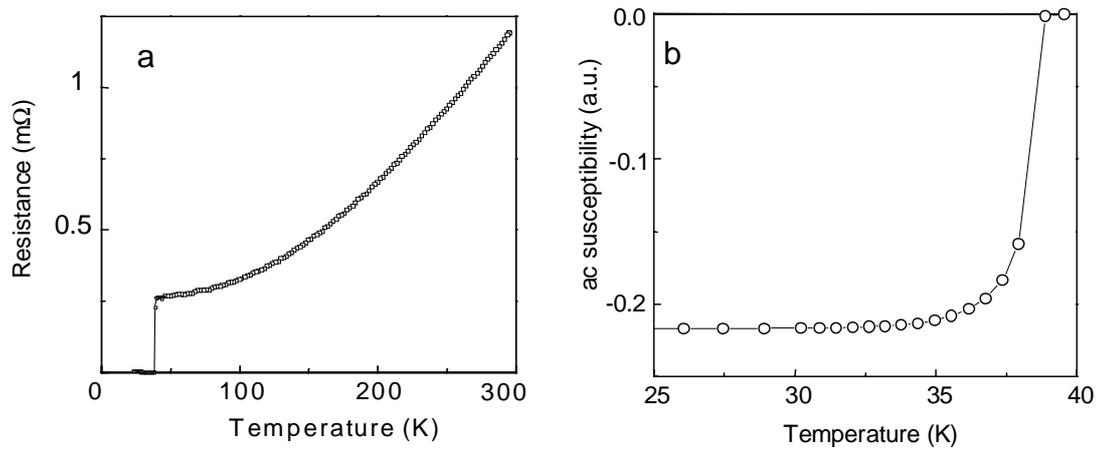

**Fig. 1**

K.-H. Müller et al., J. Alloys and Compounds

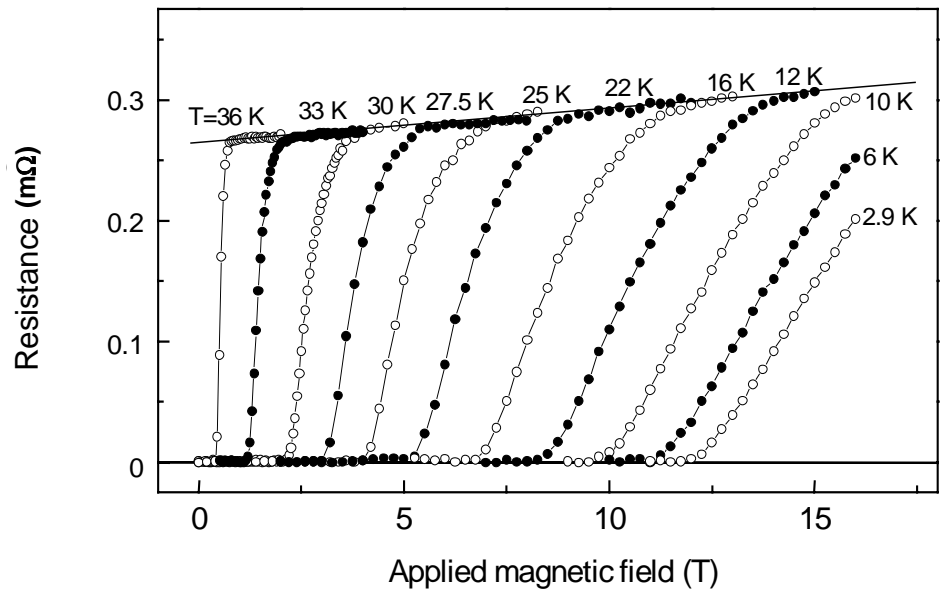

**Fig. 2**

K.-H. Müller et al., J. Alloys and Compounds

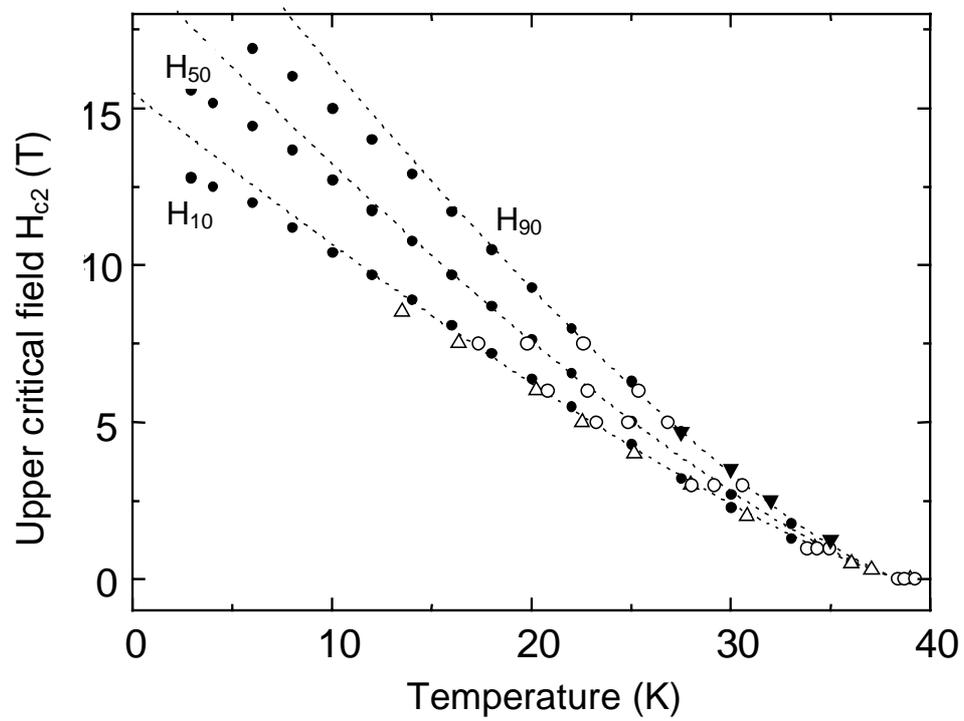

**Fig. 3**

K.-H. Müller et al., J. Alloys and Compounds

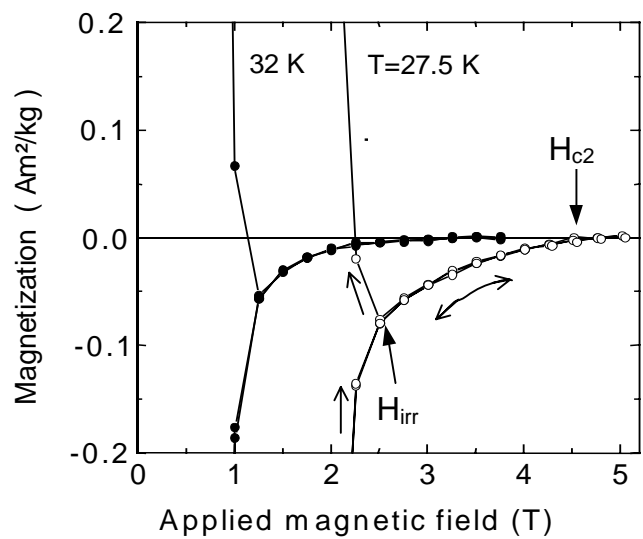

**Fig. 4**

K.-H. Müller et al., J. Alloys and Compounds